\documentclass[aps,prd,reprint,amsmath,amssymb,showpacs,superscriptaddress,nofootinbib]{revtex4-2}
\usepackage{epsfig,slashed,xcolor,bm,natbib,comment,enumitem,ulem,cancel}
\usepackage[cal=boondox]{mathalfa}

\def\nd#1#2{{\frac{d #1}{d #2}}}
\def\hnd#1#2#3{{\frac{d^{#3} #1}{d #2 ^{#3}}}}

\def\tfrac#1#2{{\textstyle\frac{#1}{#2}}}

\newcommand{\vpsi}{\varphi}

\newcommand{\comma}{, }
\newcommand{\be}{\begin{equation}}
\newcommand{\ee}{\end{equation}}
\newcommand{\bea}{\begin{eqnarray}}
\newcommand{\eea}{\end{eqnarray}}
\newcommand{\myDel}[1]{{\color{red}\ifmmode\cancel{#1}\else\sout{#1}\fi}}

\begin{document}
\title{Conformal gravity does not predict flat galaxy rotation curves}
\author{M.P.~\surname{Hobson}}
\email{mph@mrao.cam.ac.uk}
\affiliation{Astrophysics Group, Cavendish Laboratory, JJ Thomson Avenue,
Cambridge CB3 0HE\comma UK}
\author{A.N.~\surname{Lasenby}}
\email{a.n.lasenby@mrao.cam.ac.uk}
\affiliation{Astrophysics Group, Cavendish Laboratory, JJ Thomson Avenue,
Cambridge CB3 0HE\comma UK}
\affiliation{Kavli Institute for Cosmology, Madingley Road, Cambridge
  CB3 0HA, UK}

\begin{abstract}
We reconsider the widely held view that the Mannheim--Kazanas (MK)
vacuum solution for a static, spherically-symmetric system in
conformal gravity (CG) predicts flat rotation curves, such as those
observed in galaxies, without the need for dark
matter. This prediction assumes that test particles
  have fixed rest mass and follow timelike geodesics in the MK metric
  in the vacuum region exterior to a spherically-symmetric
  representation of the galactic mass distribution. Such geodesics are
  not conformally invariant, however, which leads to an apparent
  discrepancy with the analogous calculation performed in the
  conformally-equivalent Schwarzschild--de-Sitter (SdS) metric, where
  the latter does not predict flat rotation curves. This difference
  arises since the mass of particles in CG must instead be generated
  dynamically through interaction with a scalar field. The
  energy-momentum of this required scalar field means that, in a
  general conformal frame from the equivalence class of CG solutions
  outside a static, spherically-symmetric matter distribution, the
  spacetime is not given by the MK vacuum solution. A unique frame
  does exist, however, for which the metric retains the MK form, since
  the scalar field energy-momentum vanishes despite the field being
  non-zero and radially dependent. Nonetheless, we show that in both
  this MK frame and the Einstein frame, in which the scalar field is
  constant, massive particles follow timelike geodesics of the SdS
  metric, thereby resolving the apparent frame dependence of physical
  predictions and unambiguously yielding rotation curves with no flat
  region. Moreover, we show that attempts to model rising rotation
  curves by fitting the coefficient of the quadratic term in the SdS
  metric individually for each galaxy are also precluded, since the
  scalar field equation of motion introduces an additional constraint
  relative to the vacuum case, such that the coefficient of the
  quadratic term in the SdS metric is most naturally interpreted as
  proportional to a global cosmological constant.
We also comment briefly
  on how our analysis resolves the long-standing uncertainty regarding
  gravitational lensing in the MK metric.
\end{abstract}

\maketitle

\section{Introduction}
\label{sec:intro}

Conformal gravity (CG) (also known as Weyl gravity or Weyl-squared
gravity) was first proposed in 1921 by Bach \cite{Bach21}, who took
Weyl's idea of a conformally invariant gravity theory \cite{Weyl18},
but eliminated Weyl's additional vector (gauge) field, to which
Einstein had raised some theoretical objections (see \cite{SCEpaper},
however). Over the past 30 years or so, CG has attracted considerable
interest as an alternative to general relativity (GR), since it is
claimed, most notably by Mannheim and collaborators, to address
several shortcomings of GR \cite{Mannheim89,Mannheim06,Mannheim12}.

From a theoretical perspective, CG differs from GR both in incorporating the
local conformal symmetry that holds for the strong, weak and
electromagnetic interactions and in being renormalisable
\cite{Stelle77}. Conversely, whereas GR has field equations that
contain second-order derivatives of the metric and is thus unitary, CG
has fourth-order field equations and possesses a classical ghost
\cite{Riegert84b}; it is claimed that one can nonetheless construct a
unitary quantum theory by redefining its Fock space
\cite{Bender08a,Bender08b}, although this suggestion is controversial
\cite{Smilga09}.

It is from a phenomenological viewpoint, however, that CG has
generated the most interest, since it is claimed to explain various
astrophysical and cosmological observations without the need for dark
matter or dark energy. These analyses rely primarily on several exact
solutions that have been found for systems with sufficient symmetry
\cite{Riegert84,Mannheim89,Mannheim90,Mannheim91,Mannheim92,Mannheim94,Said12a,Said12b}. A
number of studies have, however, called into question many of the
claimed advantages of CG over GR, prompting a reconsideration of their
theoretical basis, most notably in the areas of cosmology,
gravitational lensing and galactic dynamics.

In a cosmological context, for homogeneous and isotropic spacetimes
the CG field equations arise solely from the energy-momentum tensor of
matter, which consists of a perfect fluid approximation both to
radiation and to a Dirac field representing `ordinary' matter, together
with a conformally coupled scalar field
\cite{Mannheim90,Mannheim01,Mannheim06}. The resulting background
cosmological evolution equations are identical to those of the
$\Lambda$CDM model derived from GR, except that the Friedmann equation
has a negative effective gravitational constant $G_{\rm eff} =
-3/(4\pi\vpsi_0^2)$, where $\vpsi_0$ is the vacuum expectation value
of the scalar field, so that isotropic radiation and matter are
repulsive, and the cosmological constant is derived from the scalar
field vacuum energy, which is proportional to $\vpsi_0^4$. This leads
to a somewhat different cosmological model to $\Lambda$CDM: the
universe is open, radiation dominates at early times to prevent a
big-bang singularity, matter is always sub-dominant, and the scalar
field dominates at late times, driving an accelerated expansion with
an effective dark energy density in the range $0 < \Omega_{\Lambda,0}
< 1$ at the current epoch, which is compatible with
observations. Indeed, it is claimed that the CG cosmology provides a
better fit to cosmological data, such as luminosity distances from
Type IA supernovae and gamma-ray bursts, and with less fine tuning
than the standard $\Lambda$CDM model, all without resort to the dark
sector
\cite{Mannheim01,Mannheim03,Mannheim11a,Nesbet11,Diaferio11,Yang13}.
The study of the growth of cosmological perturbations in CG is still
in its infancy, however, and so no predictions yet exist for the
cosmic microwave background radiation
\cite{Mannheim12,Amarasinghe19,Mannheim20}.  The CG background
cosmology does, however, already have several shortcomings. From early
on it was found to be incompatible with nucleosynthesis constraints
\cite{Knox93,Elizondo94,Lohiya99,Sethi99,Kaplinghat00}, and more
recently it has been shown to deviate significantly from high-redshift
distance moduli data derived from gamma-ray bursts and quasars,
yielding a far poorer fit than $\Lambda$CDM, and also having its own
fine-tuning and cosmic coincidence problems of a similar magnitude to
those of $\Lambda$CDM \cite{Roberts18,Lindner20}.

The majority of CG phenomenology is, however, based on the so-called
Mannheim--Kazanas (MK) vacuum solution \cite{Mannheim89} of the CG
field equations for a static, spherically-symmetric spacetime, a
solution that was, in fact, first found by Riegert \cite{Riegert84}
(note that any vacuum solution of GR, even including a cosmological
constant, is also a vacuum solution of CG, but the converse is not
true). The MK metric is compatible with solar-system tests of gravity
\cite{Wood91,Edery98,Pireaux04b,Sultana12}, but is claimed to lead to
observable differences from GR on larger scales for the trajectories
of both massless and massive particles.\footnote{It has been suggested that CG
is repulsive in the Newtonian limit and so fails solar-system tests
\cite{Flanagan06}, but it was later claimed that this conclusion is
mistaken and results from a subtlety in taking the limit in isotropic
coordinates \cite{Mannheim07}.}

Massless particles follow null geodesics, which are invariant under
conformal transformations, and have been extensively studied for the
MK metric, particularly in the context of gravitational lensing
\cite{Edery98,Pireaux04a,Pireaux04b,Amore06,Sultana10,Sultana12,Villanueva13,Cattani13,Lim17,Oguzhan19,Turner20}. Nonetheless,
the literature remains inconclusive, with different studies leading to
strongly contradictory conclusions, even regarding basic issues such
as the required sign of the linear term in the MK metric, which is key
to much of its phenomenology. These disagreements arise both from the
association of the mass of the lens with different combinations of the
parameters in the MK metric and from the choice of the geometric
definition of the deflection angle \cite{Campigotto19}; the latter is
related to the fact that the MK metric is not asymptotically flat,
which is a complication that also (in part) underlies the longstanding
confusion regarding the contribution of the cosmological constant to
gravitational lensing in the Schwarzschild--de-Sitter (SdS) metric
\cite{Rindler07,Ishak10}, a debate that has only recently been
satisfactorily resolved \cite{Butcher16}.

CG is most celebrated phenomenologically, however, for its fitting of
flat galaxy rotation curves without the need for dark matter
\cite{Mannheim93,Mannheim97,Mannheim11b,Mannheim12b,Mannheim17,OBrien18}.
These fits are based on the trajectories of massive particles in the
MK metric, with parameter values that are consistent with solar-system
tests, although the
requirement for matching the MK metric onto a static,
spherically-symmetric matter source suggests that it may not be
possible to set the parameters in a consistent manner
\cite{Perlick95,Phillips18,Negreanu18}.  More troubling, however, is that the
galaxy rotation curve analyses assume simply that massive (test)
particles follow timelike geodesics, which {\it are} affected by
conformal transformations, as is well known. Since CG is (by
construction) conformally invariant, such transformations should not
change the observable consequences of the theory, unless the conformal
symmetry is broken in some way.

A related issue is that, in any conformally invariant theory, particle
rest masses cannot be fundamental, but must instead arise dynamically,
typically through interaction with a scalar field, the mass of which
also cannot be fundamental, but arises through dynamical symmetry
breaking \cite{Dirac73}.  Thus, taking inspiration from the Standard
Model of particle physics, the matter action in CG usually includes a
Dirac field to represent `ordinary' matter, which has a Yukawa
coupling to a (conformal) scalar field with a quartic self-interaction
term \cite{Mannheim93b,Mannheim06,Edery06}.  In this
  way, one may incorporate `ordinary' matter with mass, such as that
  making up astrophysical objects such as stars and galaxies, as well
  as observers and test particles, in a way that maintains conformal
  invariance. Thus, CG is not only compatible with dynamic mass
generation, but {\it requires} it and hence also the presence of the
scalar field, without which massive (test) particles would not be
possible. Indeed, as mentioned above, the scalar field is also central
to cosmological applications of CG.

The requirement of dynamical mass generation through the presence of a
scalar field has two immediate and profound consequences for the
fitting of galaxy rotation curves with the MK metric
\cite{Wood01,Brihaye09,Horne16}. First, one cannot ignore the
energy-momentum of the scalar field and so, in general, the spacetime
outside a static, spherically-symmetric matter distribution in CG is
{\it not} given by the MK vacuum solution. Second, the mass of a test
particle depends on the value of the scalar field and hence varies
with spacetime position, in general, so the particle does {\it not}
follow a timelike geodesic unless the scalar field has the same
constant value everywhere \cite{Mannheim93b}.  It is therefore
surprising that these effects are omitted in much of the CG
literature devoted to fitting galaxy rotation curves.

Some of these issues have, however, been discussed recently in this
context. In particular, \cite{Horne16} considered a special analytical
solution of the CG field equations found in \cite{Brihaye09} for both
the metric and the scalar field in a static, spherically-symmetric
system, for which the metric still has a form equivalent to the MK
solution. Since the corresponding scalar field has a radial
dependence, however, massive (test) particles do not follow timelike
geodesics, as discussed above. Indeed, it was shown that on making a
conformal transformation to the Einstein gauge, in which the scalar
field takes a constant value everywhere and so massive particles do
follow timelike geodesics, the resulting metric is equivalent merely
to the standard SdS form, which lacks the linear term in the MK metric
that is key to the successful fitting of flat galaxy rotation
curves. This analysis was criticised in \cite{Sultana17}, however, who
pointed out that the MK metric is conformally equivalent to the SdS
metric, without the need to introduce a scalar field
\cite{Bach21,Buchdahl62,Mannheim91,Schmidt00}, and that the scalar
field in the special solution investigated in \cite{Horne16} has
vanishing energy-momentum, which was therefore considered to be
`trivial' because it has no effect on the geometry. It is unclear from
\cite{Sultana17}, however, whether these criticisms are considered to
validate previous analyses
\cite{Mannheim93,Mannheim97,Mannheim11b,Mannheim12b,Mannheim17,OBrien18} using
the MK metric to fit flat galaxy rotation curves.

In this paper, we revisit and extend the analysis in \cite{Horne16}
and address the criticisms made in \cite{Sultana17}. In particular, we
discuss how the conformal equivalence of the MK and SdS metrics, even
in the absence of a scalar field, raises concerns about the use of
timelike geodesics of the MK metric in fitting galaxy rotation
curves. Indeed, the fact that timelike geodesics of the
conformally-equivalent SdS metric are well known not to produce flat
rotation curves leads to the suspicion that the prediction in the `MK
frame' may be merely a gauge artefact. Accepting the need to include a
scalar field to facilitate dynamical mass generation, we confirm that
the scalar field in the analytical solution considered in
\cite{Horne16} has vanishing energy-momentum, as it must in order for
the MK `vacuum' solution for the metric to remain valid. Such
so-called `ghost solutions' for fields are, however, found to exist in
other physical contexts \cite{Davis75,Letelier75,Dimakis85}, and it
does not follow that the scalar field in \cite{Horne16} is dynamically
unimportant to massive particle trajectories. Indeed, we verify that
the conformal transformation required to reach the Einstein
gauge, in which the scalar field is constant everywhere
  and so its energy-momentum vanishes trivially, transforms the metric
into the SdS form, after performing a radial coordinate transformation
to recover the usual angular part of the metric. We also note that
this joint conformal and coordinate transformation is equivalent to
that previously identified in \cite{Mannheim91,Schmidt00} as
connecting the MK and SdS metrics, but merely performed in the
opposite order.  Moreover, {\it independently} of the considerations
of \cite{Brihaye09}, we show that the general form of the conformal
transformation used in \cite{Horne16} is picked out {\it uniquely} as
that which preserves the structure of {\it any} diagonal static,
spherically-symmetric metric with a radial coefficient that is (minus)
the reciprocal of its temporal one (of which both the MK and SdS
metrics are examples).
We then `close the loop' in the above considerations, by investigating
the motion of a Dirac particle with a dynamically generated mass in
the presence of a radially-dependent scalar field in a static,
spherically-symmetric spacetime. Applying this formalism to the
analytical solution considered in \cite{Horne16}, we show directly
that the equations of motion in the `MK frame' are identical to those
of timelike geodesics in the SdS metric, as they must be for conformal
invariance to hold. Hence, this both removes the dependence on
conformal frame of the predicted rotation curves that occurs in the
absence of a scalar field, and unambiguously identifies the timelike
geodesics of the SdS metric as those that are physically realised,
which do {\it not} predict flat galaxy rotation curves.  Moreover, we
show that the scalar field equation of motion introduces an additional
constraint relative to the vacuum case, such that the coefficient of
the quadratic term in the SdS metric is most naturally interpreted as
proportional to a global cosmological constant, which thus also
undermines attempts to model rising rotation curves by fitting this
coefficient separately for each galaxy, as has been considered
previously in the context of Weyl--Dirac gravity
\cite{Mirabotalebi08}. More generally, for any of type
  of galactic rotation curve usually considered (see, e.g.,
  \cite{Salucci19} for a discussion of the overall family of rotation
  curves), our analysis implies that all attempts to use the MK vacuum
  solution to avoid the need for dark matter will fail, since the real
  physical motion is just that of the conventional SdS
  metric.\footnote{This does {\it not} mean that
    conformal gravity itself cannot produce interesting dynamical
    effects, but we are concerned here only with the solution for a
    true vacuum, in which the scalar field has a vanishing
    energy-momentum tensor, and for which the MK solution can
    therefore be used.} We also briefly discuss the
  consequences of our analysis for null geodesics in the MK metric,
  and how it may be used to resolve the long-standing disagreements in
  the literature regarding gravitational lensing in CG.

Thus, in summary, the outline of our argument is as
  follows. We assume the physics to be described (everywhere) by the
  sum of the CG gravitational action (\ref{eqn:cgsg}) and the matter
  action (\ref{eqn:sipgt0lm}), which contains a Dirac field $\psi$ to
  represent `ordinary' (fermionic) matter and a scalar (compensator)
  field $\vpsi$ that enables the mass of $\psi$ to be generated
  dynamically, as required by conformal invariance. For a region with
  $\psi=0$ (apart from test particles and/or observers) outside a static,
  spherically-symmetric system, the spacetime geometry is not
  described by the MK vacuum metric in a general conformal frame from
  the equivalence class of solutions in GC, since the energy-momenum
  of the scalar field does not vanish \cite{Brihaye09}. A unique frame
  does exist, however, where the metric retains the MK form
  (\ref{eqn:MKform}--\ref{eqn:MKsoln}), since the scalar field is
  given by (\ref{eq:phisoln}), for which the energy-momentum tensor
  vanishes; the scalar field equation of motion requires that the
  additional relation (\ref{eq:constraint}) must also hold in this
  case. In both this MK frame and the Einstein frame, for which the
  scalar field is constant and the metric has the SdS form
  (\ref{eqn:SdSsoln}), massive and massless particles follow the
  timelike and null geodesics, respectively, of the SdS metric. This
  resolves the apparent frame dependence of physical predictions,
  unambiguously yields rotation curves for massive particles with no
  flat region, and resolves the long-standing debate regarding
  gravitational lensing in the MK metric.

The remainder of this paper is arranged as follows. In
Section~\ref{sec:confgrav}, we give a brief outline of conformal
gravity, including a description of its gravitational and matter
actions and the associated equations of motion. We then consider the
MK vacuum solution \cite{Riegert84,Mannheim89} for a static,
spherically-symmetric system in Section~\ref{sec:MKsoln}, and discuss
its conformal equivalence to the SdS metric without reference to any
scalar field \cite{Mannheim91,Schmidt00,Sultana17}.  In
Section~\ref{sec:rotcurvesvac}, we summarise the nature of the galaxy
rotation curves predicted in the MK and SdS vacuum solutions,
respectively.  We then discuss the necessity to introduce a scalar
field to facilitate the dynamical generation of massive (test)
particles in Section~\ref{sec:matter}, and describe the special
analytical solution of the CG field equations considered in
\cite{Brihaye09,Horne16} for both the metric and the scalar field in a
static, spherically-symmetric system. In
Section~\ref{sec:rotcurvesEinstein}, we describe the conformal
transformation of this solution to the Einstein frame and the
resulting rotation curves,
before discussing rotation curves in the MK frame directly in the
presence of a radially-varying scalar field in
Section~\ref{sec:rotcurvesMK}. We briefly comment on the implications
of our analysis for gravitational lensing in the MK metric in
Section~\ref{sec:lensing}, before concluding in
Section~\ref{sec:conc}.


\section{Conformal gravity}
\label{sec:confgrav}

Conformal gravity is interpreted geometrically in terms of a
Riemannian spacetime with metric $g_{\mu\nu}$ and has the free
gravitational action \cite{Bach21,Mannheim89,Mannheim06}
\be S_{\rm G} = \alpha \int
d^4x\,\sqrt{-g}\,C_{\rho\sigma\mu\nu}\,C^{\rho\sigma\mu\nu},
\label{eqn:cgsg}
\ee
where $\alpha$ is a dimensionless parameter and
$C_{\rho\sigma\mu\nu}$ is the Weyl tensor, which may be
written in terms of the Riemann (or curvature) tensor
$R_{\rho\sigma\mu\nu}$ and its
contractions\footnote{We adopt the following sign conventions:
  $(+,-,-,-)$ metric signature, ${R^\rho}_{\sigma\mu\nu} = 
2(\partial_{[\mu}{\Gamma^\rho}_{|\sigma|\nu]}
+{\Gamma^\rho}_{\lambda[\mu}{\Gamma^\lambda}_{|\sigma|\nu]})$, where 
the metric (Christoffel) connection ${\Gamma^\rho}_{\lambda\mu}=\tfrac{1}{2}g^{\rho\sigma}(\partial_\lambda
g_{\mu\sigma}+\partial_\mu g_{\lambda\sigma}-\partial_\sigma
g_{\lambda\mu})$, and
${R^\rho}_{\mu} = {R^{\rho\sigma}}_{\mu\sigma}$. We also
employ natural units $c=\hbar=1$ throughout,
unless otherwise stated.}
as
\bea
C_{\rho\sigma\mu\nu} &=& R_{\rho\sigma\mu\nu} \!-\!\tfrac{1}{2}(
g_{\rho\mu}R_{\sigma\nu}
\!-\!g_{\rho\nu}R_{\sigma\mu}
\!-\!g_{\sigma\mu}R_{\rho\nu}
\!+\!g_{\sigma\nu}R_{\rho\mu})\nonumber\\
&&\hspace{1.01cm} + \,\tfrac{1}{6}(g_{\rho\mu}g_{\sigma\nu}-g_{\rho\nu}g_{\sigma\mu})R.
\label{eqn:weyldef}
\eea
It is straightforward to show that under a conformal (scale)
transformation $g_{\mu\nu} \to \tilde{g}_{\mu\nu} = e^{2\rho}g_{\mu\nu}$,
where $\rho=\rho(x)$ is an arbitrary scalar function, the Weyl tensor
transforms covariantly as $\tilde{C}_{\rho\sigma\mu\nu} =
e^{-2\rho}\,C_{\rho\sigma\mu\nu}$, so that the gravitational
action $S_{\rm G}$ in (\ref{eqn:cgsg}) is invariant. Indeed, $S_{\rm
  G}$ is the unique conformally invariant action in Riemannian
spacetime. Substituting (\ref{eqn:weyldef}) into (\ref{eqn:cgsg}), one
finds that $S_{\rm G}$ may be written as
\bea
\!\!\!\!\!\!\!\!\! S_{\rm G} & = & \alpha\! \int\! d^4x\,\sqrt{-g}\,(
R_{\rho\sigma\mu\nu}R^{\rho\sigma\mu\nu}
\! -\! 2R_{\rho\sigma}R^{\rho\sigma}
\!+ \! \tfrac{1}{3}R^2) 
\label{eqn:weylsq1}\\
& = & 2\alpha\!\int\! d^4x\,\sqrt{-g}\,(R_{\rho\sigma}R^{\rho\sigma} \!-\!
\tfrac{1}{3}R^2) +\mbox{surface term},
\label{eq:equivcgaction}
\eea
where in the second line we have made use of the fact that the
Gauss--Bonnet term
$R^2-4\,R_{\rho\sigma}R^{\rho\sigma}+R_{\rho\sigma\mu\nu}R^{\rho\sigma\mu\nu}$
contributes a total derivative in $D \le 4$ dimensions.

The matter action in conformal gravity is usually taken to be
\cite{Mannheim06}
\bea
S_{\rm M}  &=&  \int d^4x\,\sqrt{-g}\,
[\tfrac{1}{2}i\bar{\psi}\gamma^\rho\,\overleftrightarrow{D}_{\!\!\rho}\psi
  - \mu\vpsi\bar{\psi}\psi \nonumber\\
&&
\qquad + \tfrac{1}{2}\nu\, (\nabla_\rho\vpsi) \,(\nabla^\rho\vpsi)
- \lambda\vpsi^4 +\tfrac{1}{12}\nu\vpsi^2R],
\label{eqn:sipgt0lm}
\eea
in which the parameters $\mu$, $\nu$ and $\lambda$ are dimensionless
and the numerical factors ensure that $S_{\rm M}$ varies only by a
surface term under a conformal transformation. In this action, $\psi$
is a Dirac field, which has Weyl weight $w=-3/2$, and the covariant
derivative in its kinetic term has the form ${D_\mu}\psi =
(\partial_\mu + \Gamma_\mu)\psi$, where the fermion spin connection
$\Gamma_\mu = \tfrac{1}{8}([\gamma^\lambda,\partial_\mu\gamma_\lambda]
- {\Gamma^\lambda}_{\nu\mu}[\gamma^\nu,\gamma_\lambda])$ and the
position-dependent quantities $\gamma_\mu = {e^a}_\mu\gamma_a$ are
related to the standard Dirac matrices ${\gamma}_a$ using the tetrad
components ${e^a}_\mu$. With a slight abuse of notation, we define
$\bar{\psi}\gamma^\rho\overleftrightarrow{D}_{\!\!\rho}\psi
\equiv \bar{\psi}\gamma^\rho D_{\rho}\psi-
(D_\rho\bar{\psi})\gamma^\rho\psi$, where the bi-directional
derivative acts only on the spinor field $\psi$ and its conjugate
$\bar{\psi}$, and not on the position-dependent gamma matrices
$\gamma^\rho$.  The (compensator) scalar field $\vpsi$, with Weyl
weight $w=-1$, has both a kinetic term and quartic self-interaction
term. The covariant derivative $\nabla_\rho$ in the former reduces to
the ordinary partial derivative, so the only direct interaction of
$\vpsi$ with the gravitational field is through its non-minimal
(conformal) coupling to the Ricci scalar. Finally, the Yukawa coupling
term between $\psi$ and $\vpsi$ is worthy of comment, since it allows
the Dirac field to acquire a mass $\mu\vpsi$ dynamically. In
particular, if one adopts the Einstein gauge $\vpsi = \vpsi_0$ (a
constant), the Dirac field has a mass $m = \mu\vpsi_0$ that is
independent of spacetime position. 

It is worth noting that the field $\vpsi$ in the
  action (\ref{eqn:sipgt0lm}) is a fundamental scalar field. In some
  of the more recent CG literature, however, the scalar field is
  instead taken to be a long-range order parameter that arises when a
  fermion bilinear associated with the Dirac field $\psi$ takes a
  non-zero expectation value $\vpsi = \langle
  0|\bar{\psi}\psi|0\rangle$ in a spontaneously broken vacuum
  $|0\rangle$ filled with negative energy fermion states
  \cite{Mannheim17}. In this case, $\vpsi$ does not appear in the
  fundamental matter action (\ref{eqn:sipgt0lm}), but an action of an
  analogous form for $\vpsi$ instead holds only within each fermion. It is then
  claimed that each fermion has its own scalar order parameter, which
  is constant outside of the fermion, where both the kinetic and Ricci
  scalar terms are absent from the action. Here, irrespective of
  the nature of the scalar field, we will confine our attention to the
  case where the total matter action has the form (\ref{eqn:sipgt0lm})
  everywhere, as proposed in \cite{Mannheim06}.

The equations of motion for the fields $g^{\mu\nu}$, $\psi$ and $\vpsi$
are obtained by varying the
total action $S_{\rm T} = S_{\rm G} + S_{\rm M} = \int
d^4x\,(\mathcal{L}_{\rm G} + \mathcal{L}_{\rm M})$ with respect to
them. On varying with
respect to $g^{\mu\nu}$ one finds that $(T_{\rm G})_{\mu\nu} + (T_{\rm
  M})_{\mu\nu} = 0$, in which
\be
 (T_{\rm G})_{\mu\nu}  \equiv
\frac{2}{\sqrt{-g}}\frac{\delta\mathcal{L}_{\rm G}}{\delta g^{\mu\nu}}
= 4\alpha(2
\nabla^\rho \nabla^\sigma + R^{\rho\sigma})C_{\mu\rho\nu\sigma},
\label{eq:bachtensor}
\ee
where the operation of the quantity in parentheses on the Weyl tensor
yields the standard expression for the Bach tensor, which is manifestly
symmetric and traceless, as expected, and scales as $e^{-4\rho}$ under
a conformal transformation (i.e.\ it has Weyl weight $w=-4$). It also
clearly contains up to fourth-order derivatives of the metric. Using
(\ref{eqn:weyldef}) and the contracted Bianchi identity $\nabla_\rho
{R^{\rho\sigma}}_{\mu\nu} -2\nabla_{[\mu} {R^{\sigma}}_{\nu]} =0$, it
is straightforward to show that any Einstein spacetime, for which 
$R_{\mu\nu} = \beta g_{\mu\nu}$ with $\beta$ a constant (and is thus a
solution of the GR vacuum field equations including a cosmological
constant), is a solution of the CG vacuum field equations.  The
converse is not true, however, since there exist solutions of $(T_{\rm
  G})_{\mu\nu}=0$ that are not Einstein spacetimes, nor conformally
equivalent to them \cite{Schmidt84,Nurowski01,Liu13}. Nonetheless, it is
worth noting that imposing a simple Neumann boundary condition on the
metric selects the vacuum solutions of GR from the
wider set of vacuum solutions of CG, thereby removing ghosts and
rendering CG and GR with a cosmological constant equivalent in a
vacuum \cite{Maldacena11}.

On including matter, its energy-momentum tensor is given by
\be
(T_{\rm M})_{\mu\nu} \equiv
\frac{2}{\sqrt{-g}}\frac{\delta\mathcal{L}_{\rm M}}{\delta g^{\mu\nu}}
= (T_\psi)_{\mu\nu} + \nu (T_\vpsi)_{\mu\nu},
\ee
where the contributions from the Dirac and scalar fields are, respectively,
\bea
(T_\psi)_{\mu\nu} &=&\tfrac{1}{2}i\bar{\psi}\gamma_{(\mu}\overleftrightarrow{D}_{\!\nu)}\psi
-g_{\mu\nu}(\tfrac{1}{2}i\bar{\psi}\gamma^\rho\overleftrightarrow{D}_{\!\!\rho}\psi
- \mu\vpsi\bar{\psi}\psi),\phantom{AA}\\
(T_\vpsi)_{\mu\nu} &=& \tfrac{1}{6}\vpsi^2 G_{\mu\nu}+\tfrac{2}{3}(\nabla_\mu\vpsi)(\nabla_\nu\vpsi)-\tfrac{1}{3}\vpsi\nabla_{\mu}\nabla_{\nu}\vpsi\nonumber\\
&&+g_{\mu\nu}\left[\tfrac{1}{3}\vpsi\,\square^2\vpsi
-\tfrac{1}{6}(\nabla_\rho\vpsi)(\nabla^\rho\vpsi)+\frac{\lambda}{\nu}\vpsi^4\right], \label{eq:tphi}
\eea
in which $\square^2 \equiv \nabla_\rho\nabla^\rho$ and $G_{\mu\nu}
\equiv R_{\mu\nu}-\tfrac{1}{2}g_{\mu\nu}R$ is the Einstein tensor. One
should note that to determine the variational derivative with respect
to $g^{\mu\nu}$ of the kinetic term for the Dirac spinor field $\psi$
in (\ref{eqn:sipgt0lm}), one must first vary with respect to the
tetrad components to obtain ${t^a}_\mu \equiv \delta\mathcal{L}_{\rm
  M}/\delta {e_a}^\mu$, from which one then has
$2\delta\mathcal{L}_{\rm M}/\delta g^{\mu\nu} =
\eta_{ab}{e^a}_{(\mu}{t^b}_{\nu)}$. It is straightforward to show that
$(T_{\rm M})_{\mu\nu}$ scales as $e^{-4\rho}$ under a conformal
transformation, like $(T_{\rm G})_{\mu\nu}$, and that its trace
vanishes by virtue of the matter equations of motion. 

The latter are obtained by varying the action with respect to the fields
$\psi$ and $\vpsi$, respectively, which yields
\bea
i\gamma^\mu D_{\mu}\psi - \mu\vpsi\psi & = & 0, \label{eq:psieq2}\\
\square^2\vpsi-\tfrac{1}{6}\vpsi R+\frac{4\lambda}{\nu}\vpsi^3
+ \frac{\mu}{\nu}\bar{\psi}\psi & = & 0.\label{eq:phieq2}
\eea
One sees that (\ref{eq:psieq2}) is the Dirac equation for a fermion field
with mass $m = \mu\vpsi$ induced by Yukawa coupling to $\vpsi$, and
(\ref{eq:phieq2}) is a Klein--Gordon equation for a massless scalar
field $\vpsi$ with a Dirac source $\mu\bar{\psi}\psi/\nu$ and a
position-dependent `Mexican hat' potential $V(\vpsi) =
-\tfrac{1}{12}R\vpsi^2 + \frac{\lambda}{\nu}\vpsi^4$.

As noted in \cite{Horne16}, by rescaling $\vpsi$ one can set the
dimensionless parameter $\nu = \pm 1$ throughout, if desired, but only
the positive value yields the `correct' sign for the kinetic energy of
the compensator scalar field. Moreover, as mentioned above, one may
use the conformal invariance of the theory to set the scalar field to
a constant $\vpsi = \vpsi_0$, which is usually termed the Einstein
gauge.


\section{Static spherically-symmetric vacuum solution}
\label{sec:MKsoln}

For any static spherically-symmetric spacetime, a particular conformal
transformation brings the line-element into the special form \cite{Mannheim89}
\be
ds^2 = B(r)\,dt^2 - \frac{dr^2}{B(r)} - r^2(d\theta^2 + \sin^2\theta\,d\phi^2).
\label{eqn:MKform}
\ee
As first shown by Riegert \cite{Riegert84} and later by Mannheim \&
Kazanas \cite{Mannheim89}, on substituting the corresponding metric
$g_{\mu\nu}$ into the vacuum CG field equations $(T_{\rm G})_{\mu\nu}
= 0$, one finds that the function $B(r)$ may be written as
\be
B(r) = 1-3\beta\gamma -\frac{\beta(2-3\beta\gamma)}{r} + \gamma r - kr^2,
\label{eqn:MKsoln}
\ee
where $\beta$, $\gamma$ and $k$ are integration
constants\footnote{Some of the CG literature, including
  \cite{Horne16}, uses the slightly different parameterisation $B(r) =
  \sqrt{1-6\beta'\gamma} -2\beta'/r + \gamma r - kr^2$, where $\beta'
  = \beta(1-\tfrac{3}{2}\beta\gamma) \approx \beta$ if the condition
  $|\beta\gamma| \ll 1$ is satisfied.}. In particular, to describe the
spacetime outside of a central mass $M$ one identifies the coefficient
of the $1/r$ term in (\ref{eqn:MKsoln}) with $-2GM/c^2$ (reinstating
$c$ for the moment), in which case $\beta \neq 0$, and the MK metric
is then in agreement with the classic solar system tests of GR
provided $|\beta\gamma| \ll 1$
\cite{Wood91,Edery98,Pireaux04b,Sultana12}. As is clear from
(\ref{eqn:MKsoln}), $\beta$ and $\gamma$ have dimensions of length and
inverse length, respectively, so that the product $\beta\gamma$ is
dimensionless\footnote{On reinstating $c$, the more physically
  relevant quantities $\beta/c^2$ and $\gamma c^2$ have dimensions of
  inverse acceleration and acceleration, respectively.}. The constant
$k$ has units of (length)$^{-2}$ and the corresponding quadratic term
$-kr^2$ in (\ref{eqn:MKsoln}) embeds the solution in a curved
background at large coordinate radius. We note that Birkhoff's theorem holds in CG, so that
  (\ref{eqn:MKsoln}) is the most general spherically-symmetric vacuum
  solution \cite{Riegert84}, which thus holds in any region where
  $(T_{\rm M})_{\mu\nu} = 0$, including exterior or interior to an
  arbitrary spherically-symmetric matter distribution.

As expected, (\ref{eqn:MKsoln}) includes the Schwarzschild solution as
a special case $(\gamma = k = 0)$. Also anticipated, but more
interesting, is that it further includes the Schwarzschild--de-Sitter
(SdS) solution $(\gamma=0)$, despite the absence of a cosmological
constant term in (\ref{eqn:cgsg}). As Bach originally showed
\cite{Bach21}, however, every static, spherically-symmetric spacetime
that is conformally related to the SdS metric is a solution of the
vacuum field equations of CG for such a system, with the converse
being proved some years later by Buchdahl \cite{Buchdahl62}. Thus, as
later verified explicitly \cite{Mannheim91,Schmidt00,Sultana17}, the
MK metric (\ref{eqn:MKsoln}) is {\it conformally equivalent} to the
SdS solution. In particular, if one redefines the radial coordinate as
$r = r'/\Omega'(r')$, where
\be
\Omega'(r') = 1-\frac{\gamma}{2-3\beta\gamma} r',
\label{eqn:conftrans1}
\ee
and then makes the conformal transformation $\tilde{g}'_{\mu\nu}(x') =
\Omega^{\prime\,2}(r')g'_{\mu\nu}(x')$, with $g'_{\mu\nu}(x') =
      {X^\rho}_\mu{X^\sigma}_\nu g_{\rho\sigma}(x(x'))$ and ${X^\rho}_\mu
      = \partial x^\rho/\partial x^{\prime\,\mu}$, one again obtains a
      line-element in the special form (\ref{eqn:MKform}), but
      expressed in terms of $r'$ and $\tilde{B}'(r')$, where
\be
\tilde{B}'(r') = 1 - \frac{\beta(2-3\beta\gamma)}{r'} - k'{r'}^2,
\label{eqn:SdSsoln}
\ee
with $k' \equiv k + \gamma^2(1-\beta\gamma)/(2-3\beta\gamma)^2$, which
has the usual SdS form that lacks the linear term present in the MK
metric (\ref{eqn:MKsoln}). One should note, however, that
  in addition to identifying the coefficient of $1/r'$ as $-2GM/c^2$,
  where $M$ is the mass of the central object, the constant $k'$ in the
  ${r'}^2$ term  may {\it also} be system dependent in CG, and need not be
  identified as $\tfrac{1}{3}\Lambda$, for some `global' cosmological
  constant $\Lambda$, which is necessary in GR.

\section{Galaxy rotation curves in a vacuum}
\label{sec:rotcurvesvac}

The modelling of galaxy rotation curves in CG is
  typically performed without invoking dark matter by using the MK
  line-element (\ref{eqn:MKform}--\ref{eqn:MKsoln}) in the (assumed)
  vacuum region exterior to a spherically-symmetric representation of
  the galactic matter distribution, and possibly also interior to a
  spherically-symmetric approximation to the matter distribution on
  much larger scales, representing the cluster or supercluster in
  which the galaxy resides and potentially extending to cosmological
  scales and including the Hubble flow
  \cite{Mannheim93,Mannheim97,Mannheim11b,Mannheim12b,Mannheim17,OBrien18}.
  In the CG literature, a somewhat complicated approach is taken to the
  description of the matter distribution interior and exterior to the
  vacuum region.  The galactic matter distribution is often considered
  as a collection of $\sim 10^{11}$ stars, which contribute terms to
  the metric coefficient $B(r)$ in the vacuum region that are
  proportional to a constant, $1/r$ and $r$, respectively, whereas the matter
  distribution on larger scales contributes a term proportional to
  $r^2$ that arises from inhomogeneities, and a further term proportional
  to $r$ that is due to the Hubble flow. Irrespective of the origins
  of these various terms, however, in the (assumed) vacuum region
  where the rotation curve is modelled, the line element must simply
  be of the general MK form (\ref{eqn:MKform}--\ref{eqn:MKsoln}),
  owing to Birkhoff's theorem in CG \cite{Riegert84}.

The conformal equivalence discussed above between the
  MK and SdS metrics then immediately raises some concerns, however,
  regarding such analyses.  In particular, a key assumption is that a
matter test particle (or star) follows a timelike geodesic in the
spacetime geometry defined by the MK line-element. This leads to the
conclusion that, for a circular orbit of coordinate radius $r$ (in the
equatorial plane $\theta = \pi/2$), the velocity $v$ of the test
particle (as measured by a stationary observer at that radius)
satisfies
\bea
v^2 &=& \frac{r^2}{B}\left(\nd{\phi}{t}\right)^2 =
\frac{r}{2B}\nd{B}{r}\nonumber \\
&=& \tfrac{1}{2} \frac{\beta(2-3\beta\gamma)r^{-1}+\gamma r -2k r^2}{
\beta(3\beta\gamma-2)r^{-1}+(1-3\beta\gamma) + \gamma r -kr^2}.
\label{eqn:MKrotcurve}
\eea

When considering a galaxy rotation curve, the weak-field limit $B
\approx 1$ holds, so that the three terms in the numerator of
(\ref{eqn:MKrotcurve}) determine its shape \cite{Horne16}.  Recalling that
$|\beta\gamma| \ll 1$, the first term in the numerator recovers the
standard Keplerian rotation curve $v^2 = \beta/r$, whereas the second
term contributes a rising component $v^2 = \tfrac{1}{2}\gamma
r$. Indeed, it is the transition between these two r\'egimes, which
occurs around $r^2 \sim 2\beta/\gamma$, that produces the
approximately flat rotation curve that resembles those observed in the
outskirts of large spiral galaxies~\cite{Rubin78}. In this context, $M
\sim 10^{11}$~M$_\odot$ and so $\beta \sim 10^{14}$~m (hence
$\beta/c^2 \sim 10^{-3}$~m$^{-1}$ s$^2$), and $\gamma$ has typically
been associated with the inverse Hubble length, such that $\gamma \sim
10^{-26}$~m$^{-1}$ (hence $\gamma c^2 \sim 10^{-9}$~m s$^{-2}$). Thus,
$|\beta\gamma| \sim 10^{-12}$, which amply satisfies the requirement
$|\beta\gamma| \ll 1$. With these values of $\beta$ and $\gamma$, the
two contributions to the overall rotation curve are of a similar
magnitude for $r \sim 10^{20}$ m or $\sim 5$~kpc, which corresponds
roughly to the size of a galaxy.  Some refinement of the model is
necessary to accommodate the rising rotation curves observed in
smaller dwarf galaxies, so more recent analyses assume $\gamma(M) =
\gamma_0(1+M/M_0)$, where $\gamma_0 \sim 10^{-28}$~m$^{-1}$ and $M_0
\sim 10^{10}$ M$_\odot$~\cite{Mannheim93,Mannheim97}, such that
$\gamma \sim 10^{-27}$~m$^{-1}$ for $M \sim 10^{11}$~M$_\odot$ and the
transition between r\'egimes occurs at $r \sim 15$~kpc. Moreover, to
model the flat rotation curves in the outskirts of particularly large
galaxies \cite{Mannheim12}, one requires $k > 0$ so that the falling
quadratic term $-kr^2$ counters the rising term $\gamma r$ in the
numerator of (\ref{eqn:MKsoln}). A reasonable fit is obtained if $k
\sim 10^{-49}$~m$^{-2}$ $\sim (100~\mbox{Mpc})^{-2}$, where 100~Mpc coincides
with the typical size of structures in the cosmic web; this also serves
to eliminate bound circular orbits beyond the `watershed' radius $r =
|\gamma/2k| \sim 150$~kpc.

In any case, it is clear that the linear term $\gamma r$ in
(\ref{eqn:MKsoln}) is crucial for producing flat rotation curves that
resemble those observed in galaxies. It is therefore concerning that
the MK metric is conformally equivalent to the SdS form
(\ref{eqn:SdSsoln}), for which the linear term is absent. If one again
assumes simply that a matter test particle follows a timelike geodesic
in the spacetime geometry, but now that defined by the SdS
line-element (\ref{eqn:SdSsoln}), the corresponding circular velocity
$\tilde{v}'$ of the test particle satisfies
\be
\tilde{v}^{\prime 2}  
= \frac{r'}{2\tilde{B}'}\nd{\tilde{B}'}{r'}
 =  \tfrac{1}{2} \frac{\beta(2-3\beta\gamma)r^{\prime\,-1}-2k' r^{\prime\,2}}{
\beta(3\beta\gamma-2)r^{\prime\,-1}+1-k'r^{\prime\,2}}.
\label{eqn:SdSrotcurve}
\ee
By
analogy with the argument given above, $\tilde{B}' \approx 1$ in this
astrophysical application, so that the rotation curve is determined by
the two terms in the numerator of (\ref{eqn:SdSrotcurve}). Again
assuming $|\beta\gamma| \ll 1$, the first term similarly recovers the
standard Keplerian result $\tilde{v}^{\prime 2} = \beta/r'$, and the
second term contributes $\tilde{v}^{\prime 2} = -k'r^{\prime\,2}$,
where $k' \approx k + \tfrac{1}{4}\gamma^2$. Moreover, assuming
similar values for $\beta$, $\gamma$ and $k$ as used above, then $k'
\approx k$ and the rotation curve falls for all values of $r'$ until
bound circular orbits are eliminated beyond the new watershed radius
$r' = |\beta/k|^{1/3} \sim 20$~kpc. Thus, in the `SdS frame', there is
no region with a flat rotation curve, as
expected.\footnote{The SdS metric
  has nonetheless been used to model a small number of spiral galaxies with
  rotation curves that are rising for all observed values of $r'$, by
  fitting a {\it negative} value of $k'$ {\it separately} for each
  galaxy, albeit in the context of Weyl--Dirac gravity \cite{Mirabotalebi08}.}

Since the transformations linking the two metrics (\ref{eqn:MKsoln})
and (\ref{eqn:SdSsoln}) leave the CG gravitational action
(\ref{eqn:cgsg}) invariant, however, they should not change the
observable physical consequences of the theory, unless conformal
invariance is broken in some way. The ambiguity in the predicted
rotation curves arises from the assumption that a test particle of
fixed rest mass $m$ follows a timelike geodesic, which is based on the
standard postulate in GR that the worldline extremises the particle
action $S_{\rm p} = -m \int ds$, where $ds$ is the spacetime interval.
This action is unsuitable in CG, however, since it is not invariant
under conformal transformations \cite{Wood91}.  This leads to the
suspicion that the flat rotation curves predicted in the `MK frame'
may be merely a gauge artefact.


\section{Dynamic generation of test particle masses}
\label{sec:matter}

The question then naturally arises as to which of the rotation curves
(\ref{eqn:MKrotcurve}) or (\ref{eqn:SdSrotcurve}), if either, is
physically realised. The key to resolving this ambiguity is to
recognise that the (massive) test particle in either scenario
represents some form of `ordinary' matter, typically described by a
Dirac field. Thus, even when using the MK and SdS metrics, which both
satisfy the vacuum field equations of CG, one must still consider how
to introduce matter in the form of a Dirac field into the theory in a
consistent manner, in order to model correctly the trajectories of
massive test particles in the vacuum region.

As discussed in Section~\ref{sec:confgrav}, the appropriate form for
the matter action in CG has the form (\ref{eqn:sipgt0lm}).  In
particular, to satisfy conformal invariance, the Dirac field must acquire a
mass dynamically through the Yukawa coupling term
$\mu\vpsi\bar{\psi}\psi$, which thus {\it necessitates} the
introduction of a scalar (compensator) field $\vpsi$ that is non-zero
everywhere (except perhaps at infinity). If the Dirac field $\psi$
represents only the test particle, then one need solve only the
coupled equations of motion of the metric $g^{\mu\nu}$ and scalar
field $\vpsi$: the former is given by $(T_{\rm G})_{\mu\nu} +
\nu(T_\vpsi)_{\mu\nu} = 0$ using (\ref{eq:bachtensor}) and
(\ref{eq:tphi}), and the latter by (\ref{eq:phieq2}) with $\psi=0$.

The solutions of these equations for a static spherically-symmetric
spacetime are investigated in \cite{Brihaye09}. In general, the
non-zero energy-momentum of the scalar field means
that the metric does not have the MK form. A special analytical
solution is identified, however, for which the scalar field is
everywhere non-zero and finite, given by
\be
\vpsi(r) = \vpsi_0\left(1+\frac{r}{a}\right)^{-1}, 
\label{eq:phisoln}
\ee
where $\vpsi_0$ and $a$ are finite positive
constants, but its entire energy-momentum tensor
nonetheless vanishes identically, $(T_\vpsi)_{\mu\nu}
= 0$ \cite{Brihaye09,Horne16,Sultana17}.  In this case, the
metric then clearly still satisfies the CG vacuum field equations
$(T_{\rm G})_{\mu\nu}=0$ and so can be written in the form of the MK
line-element (\ref{eqn:MKform}--\ref{eqn:MKsoln}), in which case $a =
(2-3\beta\gamma)/\gamma$ in (\ref{eq:phisoln}). One should also note
that the scalar field equation of motion (\ref{eq:phieq2}), with
$\psi=0$, introduces an {\it additional} constraint relative to the
vacuum case, which imposes the following relationship between the
integration constants in the MK metric:
\be 
k + \frac{\gamma^2(1-\beta\gamma)}{(2-3\beta\gamma)^2} =
-2\lambda\vpsi_0^2. 
\label{eq:constraint}
\ee
Thus, the constant $a$ in (\ref{eq:phisoln}) is
  expressible wholly in terms of the coefficients in the MK metric
  (\ref{eqn:MKform}), and the overall normalisation $\vpsi_0$ in
  (\ref{eq:phisoln}) may also be expressed in terms of these
  coefficients and the constant $\lambda$ appearing the scalar field
  potential energy term in (\ref{eqn:sipgt0lm}). Consequently, one is
  {\it not} free to assume different values for the constants $a$ and
  $\vpsi_0$, as is done in some of the CG literature, where it is
  assumed that the $\vpsi$ field in which a test particle (star) moves
  is somehow `generated' by the test particle itself, rather
  than being an `ambient' field that permeates the vacuum region
  \cite{Mannheim12b}.

For metrics of the special form (\ref{eqn:MKform}), one may show that
(\ref{eq:phisoln}) is the most general static, spherically-symmetric
form for $\vpsi$ that is a so-called `ghost solution', i.e.\ a
non-zero matter field configuration that solves the equations of
motion but has vanishing energy-momentum.\footnote{It is worth noting
  that other `ghost solutions' exist, for example for the Dirac field
  in both Einstein--Weyl and Einstein--Cartan gravity for certain
  systems \cite{Davis75,Letelier75,Dimakis85}.}  Thus, if the
line-element has the general MK form
(\ref{eqn:MKform}--\ref{eqn:MKsoln}), the energy-momentum tensor
(\ref{eq:tphi}) of the scalar field vanishes {\it if and only if}
$\vpsi(r)$ has the form (\ref{eq:phisoln}). In particular, it does
{\it not} vanish for a constant scalar field
$\vpsi(r)=\vpsi_0$. Hence, irrespective of the assumed physical nature
of the scalar field, provided the matter action has
  the form (\ref{eqn:sipgt0lm}), $\vpsi$ {\it cannot} have a constant
value if the spacetime geometry is described by the general MK metric,
since this combination is prohibited by the vacuum field equations. If
one sets $\vpsi(r) = \vpsi_0$, the scalar field energy-momentum tensor
vanishes only if $G_{\mu\nu} + 6\lambda\vpsi_0^2g_{\mu\nu} = 0$, so
that the only vacuum metric allowed has the SdS form
(\ref{eqn:SdSsoln}) with $k' = -2\lambda\vpsi_0^2$. These
considerations cast doubt on much of the CG literature devoted to the
fitting of galaxy rotation curves
\cite{Mannheim93,Mannheim97,Mannheim11b,Mannheim12b,Mannheim17,OBrien18}.

As we discuss in the next section, the form
  (\ref{eq:phisoln}) is determined directly by the required form of
  the transformation to the Einstein gauge $\vpsi(r) = \vpsi_0$ for
  metrics of the special form (\ref{eqn:MKform}). Since the scalar
  field energy-momentum vanishes straightforwardly in the Einstein
  gauge, one may view the form (\ref{eq:phisoln}) as merely an
  artefact of solving the equations of motion in a gauge in which the
  metric is assumed to have the special form
  (\ref{eqn:MKform}).

In any case, the immediate consequence of
(\ref{eq:phisoln}) is that a (fermionic) test particle has a
dynamically induced mass $m=\mu\vpsi$ that varies with coordinate
radius and hence it does not follow a timelike geodesic, which
violates the key assumption made in deriving the rotation curve
(\ref{eqn:MKrotcurve}) in the MK frame.

\section{Galaxy rotation curves in the Einstein frame}
\label{sec:rotcurvesEinstein}

Rather than including the effect of the radially-dependent scalar
field on the massive test particle trajectory directly in the MK
frame, we first consider the approach used in \cite{Horne16}, where
one takes advantage of the conformal invariance of the theory and
performs a conformal transformation to the Einstein frame, in which
the scalar field has the constant value $\vpsi_0$ and so the test
particle has the same mass $m=\mu\vpsi_0$ everywhere and hence does
follow a timelike geodesic.  

Instead of considering the MK metric directly, however, it is more
informative to illustrate a conformal transformation to the Einstein
gauge for a {\it general} static, spherically-symmetric
metric of the form
\be
ds^2 = A(r)\,dt^2 - \frac{dr^2}{B(r)} - r^2(d\theta^2 + \sin^2\theta\,d\phi^2),
\label{eqn:genform}
\ee
which clearly coincides with the special form (\ref{eqn:MKform}) when
$A(r) = B(r)$; we also consider a general form for $\vpsi(r)$. Since
$\vpsi$ has Weyl weight $w=-1$, under a (radial) conformal
transformation one has $\vpsi(r) \to \tilde{\vpsi}(r) =
\vpsi(r)/\Omega(r)$, and so the required conformal transformation to
achieve $\tilde{\vpsi}(r)=\vpsi_0$ is simply $\Omega(r) =
\vpsi(r)/\vpsi_0$, and the metric becomes $\tilde{g}_{\mu\nu}(x) =
\Omega^2(r) g_{\mu\nu}(x)$. To bring the angular part of the metric
back into the standard form in (\ref{eqn:genform}), but expressed in
terms of a new radial coordinate $r'$, one must then perform the
(radial) coordinate transformation $r' = r\Omega(r)$ to obtain
$\tilde{g}'_{\mu\nu}(x') = {X^\rho}_\mu{X^\sigma}_\nu
\tilde{g}_{\rho\sigma}(x(x'))$. In so doing, one finds that the
resulting line-element again has the form (\ref{eqn:genform}), but
expressed in terms of the new radial coordinate $r'$ and the metric
functions
\be
\tilde{A}'(r') = \Omega^2(r(r')) A(r(r')), \quad
\tilde{B}'(r') = f^2(r(r')) B(r(r')),
\ee
where we have defined the function $f(r) \equiv
1+r\frac{d\ln\Omega(r)}{dr}$.  Thus, as one might expect, even if
(\ref{eqn:genform}) has the special form (\ref{eqn:MKform}) in which
$A(r) = B(r)$, this form is {\it not} preserved in
  general by these transformations. Indeed, this is achieved {\it
  only} if $f^2(r) = \Omega^2(r)$, which is readily solved on
demanding that $\Omega(r)$ is finite everywhere to obtain $\Omega(r) =
\vpsi(r)/\vpsi_0$, where $\vpsi(r)$ is given by (\ref{eq:phisoln})
with $a$ arbitrary. Thus, {\it independently} of the considerations in
\cite{Brihaye09}, the general form (\ref{eq:phisoln}) for the scalar
field is picked out {\it uniquely} as that for which the corresponding
(finite) conformal transformation to the Einstein frame preserves the
special form (\ref{eqn:MKform}) of the metric. Equally, starting from
a metric of the special form (\ref{eqn:MKform}), if $\vpsi(r)$ does
not have the form (\ref{eq:phisoln}), where $a$ may be arbitrary, then
the resulting transformed line-element in the Einstein frame does {\it
  not} also have this special form.

Adopting the form (\ref{eq:phisoln}) for the scalar field and applying
the above approach to the MK metric, for which $B(r)$ is given by
(\ref{eqn:MKsoln}) and $a = (2-3\beta\gamma)/\gamma$, the required
conformal transformation to the Einstein gauge is simply
\be
\Omega(r) = \left(1 + \frac{\gamma}{2-3\beta\gamma}r\right)^{-1},
\label{eqn:conftrans2}
\ee
and one finds that $\tilde{B}'(r')$ is again given precisely by the
SdS form (\ref{eqn:SdSsoln}). As might be expected by comparing the
forms of the conformal transformations (\ref{eqn:conftrans1}) and
(\ref{eqn:conftrans2}), the coordinate $r'$ has the same form as that
originally used in \cite{Mannheim91,Schmidt00,Sultana17} to obtain
(\ref{eqn:SdSsoln}), which is given by $r' =
r(1+\frac{\gamma}{2-3\beta\gamma}r)^{-1}$. Note that this tends to the
finite value $r' \to (2-3\beta\gamma)/\gamma$ as $r \to
\infty$. Indeed, the only difference in the two approaches is that the
conformal and (radial) coordinate transformations are performed in
opposite orders.

In any case, since a test particle has a constant mass in the Einstein
gauge, and thus follows a timelike geodesic, one thus identifies
(\ref{eqn:SdSrotcurve}) as the rotation curve that is physically
realised. Moreover, one should note from (\ref{eqn:SdSsoln}) and
(\ref{eq:constraint}) that one now requires
$k'=-2\lambda\vpsi_0^2$ in
(\ref{eqn:SdSrotcurve}). If one assumes that $\vpsi_0$ in
(\ref{eq:phisoln}) may be system dependent, then there remains the
possibility that exists in the vacuum case of attempting to model some
(typically rising) rotation curves by fitting for $k'$ separately for
each galaxy, as in \cite{Mirabotalebi08}. Such an assumption seems
questionable when viewed in the Einstein gauge, however, where
$\vpsi_0$ is more naturally interpreted as a system-independent
quantity that leads to a `global' cosmological constant $\Lambda =
-6\lambda\vpsi_0^2$. In this case, one may therefore
no longer fit for $k'$ separately for each galaxy, or at all if one
considers $\Lambda$ to be fixed by cosmological
observations. It is also worth noting that, to obtain
  a positive cosmological constant $\Lambda$, one must have $\lambda
  <0$, which thus requires a negative scalar field vacuum energy
  $\lambda\vpsi^4_0$, at least with the usual sign conventions adopted
  in the matter action (\ref{eqn:sipgt0lm}).

\section{Galaxy rotation curves in the MK frame}
\label{sec:rotcurvesMK}

We now `close the loop' in our considerations by instead including the
effect of a radially-dependent scalar field on massive particle
trajectories directly in the MK frame. In the interests of generality,
however, we will first present our results for an arbitrary static,
spherically-symmetric metric of the special form (\ref{eqn:MKform})
and an arbitrary radial scalar field $\vpsi(r)$, before explicitly
considering the case of the MK metric (\ref{eqn:MKsoln}) and the
scalar field configuration (\ref{eq:phisoln}).\footnote{It is, in
  fact, straightforward to perform the calculation for a metric of the
  {\it general} form (\ref{eqn:genform}) and an arbitrary radial
  scalar field $\vpsi(r)$, but we present here only the results for
  $A(r)=B(r)$ for the sake of brevity. We discuss the wider
  implications of the gauge choice $A(r)=B(r)$ in \cite{ANL-MPH}, and
  in particular describe the manner in which it distorts the scaling
  properties of variables, thereby making it extremely difficult to
  identify `intrinsic' $\vpsi$-independent quantities that may be used
  for performing all calculations, including the derivation of the
  geodesic equations.}

We begin by again assuming that a matter
test particle is represented by a Dirac field, and construct an
appropriate action from which its equation of motion can be
derived. The construction of the action for a spin-1/2 point particle
and the subsequent transition to the full classical approximation in
which the particle spin is then neglected is discussed in
\cite{eWGTpaper}. In the presence of a Yukawa coupling to a scalar
compensator field, this yields
\begin{equation}
S_{\rm p} = - \int d\zeta\, [p_a u^{a}-\tfrac{1}{2}e(p_a p^a - \mu^2\vpsi^2)],
\label{WGTppaction}
\end{equation}
where the dynamical variables are the tetrad components of the
particle 4-momentum $p_a(\zeta) = {e_a}^\mu p_\mu(\zeta)$ and
4-velocity $u^a(\zeta) = {e^a}_\mu dx^\mu(\zeta)/d\zeta$, and
the einbein $e(\zeta)$ along the worldline $x^\mu(\zeta)$, which
is parameterised by $\zeta$. 

As also shown in \cite{eWGTpaper}, in order that $u^au_a
  = u^\mu u_\mu = 1$ for a massive particle, the einbein must take the
  form $e=1/(\mu\vpsi)$.  In this case, the Weyl weights of the
  quantities appearing in (\ref{WGTppaction}) are $w(p_a)=-1$,
  $w(u^a)=0$, $w(e)=1$, $w(\zeta)=1$ and $w(\vpsi)=-1$, so that the
  action is indeed scale-invariant. On varying the action with respect
  to the dynamical variables $p_a$, $x^\mu$ and $e$, one finds that
  the particle equation of motion may then be written in the
  coordinate frame as
\be
u^\nu \nabla_\nu u^\mu = (g^{\mu\nu}-u^\mu u^\nu)\partial_\nu \ln\vpsi.
\label{eqn:confeom}
\ee
Thus, as expected, it is only when $\vpsi$ is constant that the
particle follows a timelike geodesic. It is worth noting that one may
also arrive at the equation of motion (\ref{eqn:confeom}) in a more
heuristic manner by simply positing the action of a particle with
position-dependent mass $m(x)=\mu\vpsi(x)$ to be $S_{\rm p} = -\int
m(x)\,ds = -\mu\int\vpsi(x)\,ds$, which is a straightforward
generalisation of the usual particle action assumed in GR
\cite{Wood91}, and identifying the parameters $\zeta$ and $s$
(although we shall draw a distinction between $\zeta$ and proper
time below).

Assuming a static, spherically-symmetric system with $\vpsi=\vpsi(r)$
and a line-element in the special form (\ref{eqn:MKform}), one finds
that for a massive particle worldline $x^\mu(\zeta)$ in the
equatorial plane $\theta = \pi/2$, the $t$- and $\phi$-equations of
motion are 
\be
B\Omega\nd{t}{\zeta} = \mathcal{k},\qquad
r^2\Omega\nd{\phi}{\zeta} = \mathcal{h},
\label{eqn:parteoms1}
\ee
where $\mathcal{k}$ and $\mathcal{h}$ are constants, and
we may replace the $r$-equation of motion with the much simpler first
integral $u^\mu u_\mu= 1$, which reads
\be
B\left(\nd{t}{\zeta}\right)^2 -B^{-1}\left(\nd{r}{\zeta}\right)^2 
- r^2\left(\nd{\phi}{\zeta}\right)^2 = 1,
\label{eqn:parteoms2}
\ee
where $\Omega(r) \equiv \vpsi(r)/\vpsi_0$ and the constants
$\mathcal{k}$ and $\mathcal{h}$ are
defined such that one recovers the familiar timelike geodesic
equations in GR for an affine parameter $\zeta$ if $\vpsi(r) =
\vpsi_0$ and so $\Omega =1$.

As discussed in \cite{SCEpaper}, however, the parameter $\zeta$
cannot be interpreted as the proper time of the particle, since it has
Weyl weight $w(\zeta) = 1$ and so it is not invariant under
conformal transformations. Rather, the proper time interval measured
by some (atomic) clock moving with the particle is instead given by
$d\tau \propto \vpsi\,d\zeta$, which is correctly invariant under
conformal transformations. Without loss of generality, one may choose
the constant of proportionality such that $d\tau =
(\vpsi/\vpsi_0)\,d\zeta = \Omega\,d\zeta$, and so $d\tau$ and
$d\zeta$ coincide if $\vpsi(r) = \vpsi_0$. Thus, when expressed in
terms of the proper time $\tau$ of the particle, and denoting
$d/d\tau$ by an overdot, the equations of motion
(\ref{eqn:parteoms1}--\ref{eqn:parteoms2}) become
\be
B\Omega^2\dot{t} = \mathcal{k},\quad
r^2\Omega^2\dot{\phi} = \mathcal{h},\quad
B\dot{t}^2 -B^{-1}\dot{r}^2 
- r^2\dot{\phi}^2 = \Omega^{-2}.
\label{eqn:parteoms3}
\ee

On substituting the first two equations
into the third, one straightforwardly obtains the `energy' equation
for massive particle trajectories,
\be
\dot{r}^2\Omega^4 + \left(\Omega^2+\frac{\mathcal{h}^2}{r^2}\right)B 
= \mathcal{k}^2.
\label{eqn:energy}
\ee
Then substituting $\dot{r} = \dot{\phi}\,dr/d\phi =
(\mathcal{h}/r^2\Omega^2)\,dr/d\phi$, defining the reciprocal radial variable $u
\equiv 1/r$ and differentiating with respect to $\phi$, one obtains
the `shape' (or `orbit') equation for massive particle trajectories,
\be 
\hnd{u}{\phi}{2} + \tfrac{1}{2}\nd{}{u}\left[\left(u^2 +
  \frac{\hat{\Omega}^2}{\mathcal{h}^2}\right)\hat{B}\right] = 0,
\label{eqn:shape}
\ee
where we have defined the functions $\hat{B}(u) \equiv B(1/u)$ and
$\hat{\Omega}(u) \equiv \Omega(1/u)$. As expected, if $\Omega=1$ the
equations (\ref{eqn:parteoms3}--\ref{eqn:shape}) reduce to the
familiar equations for a timelike geodesic in the equatorial
plane $\theta=\pi/2$ of a static, spherically-symmetric system
with line-element of the special form (\ref{eqn:MKform})
\cite{GRbook}.

We now specialise to the case where the scalar field $\vpsi(r)$ has
the form (\ref{eq:phisoln}), with $a$ arbitrary, for which
$(r\Omega)\,\dot{} = \dot{r}\Omega^2$. In this case, on defining the
new radial coordinate $r' = r\Omega(r)$ and metric function
$\tilde{B}'(r') = B(r(r'))\Omega^2(r(r'))$, a
dramatic simplification takes place whereby the
equations of motion (\ref{eqn:parteoms3}) become
\be
\tilde{B}'\dot{t} = \mathcal{k},\quad
r^{\prime\,2}\dot{\phi} = \mathcal{h},\quad
\tilde{B}'\dot{t}^2 -\tilde{B}^{\prime\,-1}\dot{r}^{\prime\,2} 
- r^{\prime\,2}\dot{\phi}^2 = 1,
\label{eqn:parteoms4}
\ee
and the `energy' and `shape' equations (\ref{eqn:energy}) and
(\ref{eqn:shape}), respectively, have the forms
\bea
\dot{r}^{\prime\,2} + \left(1+\frac{\mathcal{h}^2}{r^{\prime\,2}}\right)\tilde{B}'
& = & \mathcal{k}^2, \label{eq:massiveenergy}\\
\hnd{u'}{\phi}{2} + \tfrac{1}{2}\nd{}{u'}\left[\left(u^{\prime
    2} +
  \frac{1}{\mathcal{h}^2}\right)\hat{B}'\right] & = & 0,
\label{eq:massiveshape}
\eea
where $u' = 1/r'$ and $\hat{B}'(u') = \tilde{B}'(1/u')$. These
equations have precisely the same form in terms of the particle proper time
as the equations for a timelike geodesic in the equatorial plane
$\theta=\pi/2$ of a static, spherically-symmetric spacetime with
line-element of the special form (\ref{eqn:MKform}), but in terms of
the radial variable $r'$ and the metric function $\tilde{B}'(r')$.

Specialising further to the case where $B(r)$ has the MK form
(\ref{eqn:MKsoln}) and $a = (2-3\beta\gamma)/\gamma$, so that
$\Omega(r)$ is given by (\ref{eqn:conftrans2}), we know from
Section~\ref{sec:rotcurvesEinstein} that $\tilde{B}'(r')$, as defined
above, has the SdS form (\ref{eqn:SdSsoln}) with $k' =
-2\lambda\vpsi_0^2$.  Thus, by explicitly taking into
account the presence of the radially-dependent scalar field in the MK
frame, we have arrived directly at the same conclusion that we reached
previously in Section~\ref{sec:rotcurvesEinstein} by transforming to
the Einstein frame, namely that in terms of the radial coordinate $r'$
it is the rotation curve (\ref{eqn:SdSrotcurve}) that is physically
realised, which does not have the flat region observed in
galaxies. This thereby eliminates, as it must, the ambiguity discussed
in Section~\ref{sec:rotcurvesvac}, where the physical predictions
appeared to depend on the conformal frame in which the calculation was
performed.

\section{Gravitational lensing}
\label{sec:lensing}

As discussed in Section~\ref{sec:intro}, the literature concerning
gravitational lensing in the MK metric remains inconclusive, with
considerable disagreement between different studies
\cite{Edery98,Pireaux04a,Pireaux04b,Amore06,Sultana10,Sultana12,Villanueva13,Cattani13,Lim17,Oguzhan19,Turner20}. It
is uncontroversial, however, that null geodesics are unaffected by
conformal transformations. Thus, on performing a joint conformal and
(radial) coordinate transformation, such as those considered above, if
$r=r(\phi)$ is the original orbit equation for a massless particle in
the equatorial plane $\theta=\pi/2$, then it is transformed simply
into $r' = r'(r(\phi))$. Although this, of course, leads to local
changes in the trajectory, it has been suggested previously that the
global behaviour is unaffected and, in particular, that the range of
$\phi$ in the orbit equation remains unchanged, and hence so too does
the scattering angle or deflection \cite{Edery01}. This is valid,
however, {\it only} if $r' \to r$ as $r \to \infty$, which does not
hold for the transformation from the MK frame to the Einstein frame
discussed in Section~\ref{sec:rotcurvesEinstein}, for which $r' \to
(2-3\beta\gamma)/\gamma$ as $r \to \infty$. Consequently, the
scattering angle or deflection will, in general, {\it differ} between
the two frames.  This does not correspond to any {\it physical}
difference, however, but is merely a consequence of the fact that $r'$
is finite as $r \to \infty$ for the two radial coordinates
used\footnote{For the same reason, the scattering angle or deflection
  of {\it massive} particle trajectories will also differ between the
  MK and Einstein frames, even after taking into account the effect of
  the scalar field.}.

Indeed, it is instructive in this context to revisit the
  calculation of particle trajectories in the MK frame, but for the case of
  massless particles.  As discussed in \cite{eWGTpaper}, one may
  deduce the motion of photons by directly considering the dynamics of
  the electromagnetic field in the gravitational background, but one
  may also arrive at the same conclusions by reconsidering the
  particle action (\ref{WGTppaction}), which is immediately applicable
  to massless fermions (such as a neutrino) by setting $\mu=0$. In
  this case, one finds that $u^\mu u_\mu = 0$ irrespective of the form
  of the einbein $e$, which one is therefore free to choose in the
  most convenient manner. Here we take
  $e=1/\vpsi$, so that the weight of each variable in the action matches
  that in the massive case discussed in
  Section~\ref{sec:rotcurvesMK}.

One then finds that the equation of motion (\ref{eqn:confeom}) is
replaced by $u^\nu \nabla_\nu u^\mu=-u^\mu u^\nu\partial_\nu
\ln\vpsi$, but this nonetheless leaves the $t$- and $\phi$-equations
of motion (\ref{eqn:parteoms1}) unchanged and the first integral
(\ref{eqn:parteoms2}) differs only in that the right-hand side is
zero. Once again, one cannot use $\zeta$ to parameterise the particle
trajectory since it has Weyl weight $w(\zeta)=1$ and so is not
invariant under conformal transformations. As previously, the
appropriate invariant measure is $d\tau = (\vpsi/\vpsi_0)\,d\zeta =
\Omega\,d\zeta$, although $\tau$ cannot be interpreted as a proper
time in this case, since the worldline is null\footnote{For arbitrary
  $e$, the invariant interval is proportional to $\vpsi^2
  e\,d\zeta$.}. Following through an analogous calculation to that
performed in Section~\ref{sec:rotcurvesMK}, one finds that the
`energy' and `shape' equations corresponding to
(\ref{eq:massiveenergy}) and (\ref{eq:massiveshape}) are
\bea
\dot{r}^{\prime\,2} + \frac{\mathcal{h}^2}{r^{\prime\,2}}\tilde{B}'
& = & \mathcal{k}^2, \label{eq:masslessenergy}\\
\hnd{u'}{\phi}{2} + \tfrac{1}{2}\nd{}{u'}\left(u^{\prime
    2}\hat{B}'\right) & = & 0,
\label{eq:masslessshape}
\eea
which have the same form as the equations in terms of an
affine parameter for a null geodesic in the equatorial plane
$\theta=\pi/2$ of a spacetime with
line-element of the special form (\ref{eqn:MKform}), but in terms of
the new radial coordinate $r'=1/u'=r\Omega(r)$ and 
$\tilde{B}'(r') = B(r(r'))\Omega^2(r(r'))$.

As in Section~\ref{sec:rotcurvesMK}, if one now specialises to the
case where the original metric function $B(r)$ has the MK form
(\ref{eqn:MKsoln}) and $a = (2-3\beta\gamma)/\gamma$, so that
$\Omega(r)$ is given by (\ref{eqn:conftrans2}), then the function
$\tilde{B}'(r')$ has the SdS form (\ref{eqn:SdSsoln}) with $k' =
-2\lambda\vpsi_0^2$. Thus, in terms of the radial coordinate
$r'$, the trajectories of massless particles in the MK frame follow
null geodesics of the SdS metric.  This therefore resolves the
uncertainty in the literature regarding gravitational lensing in the
MK frame, since the SdS metric lacks the linear term that has prompted
so much debate in the CG literature, and the consequences of the
quadratic `cosmological constant' term have recently been properly
determined \cite{Butcher16}.  Hence,
one may arrive at unambiguous predictions for gravitational lensing
that can then be easily recast in terms of the original radial coordinate $r$
used in the MK form for $B(r)$ in (\ref{eqn:MKsoln}), if
desired.

Finally, as shown in Section~\ref{sec:rotcurvesEinstein}, we note
again here that one may only reach another metric having the special
form (\ref{eqn:MKform}) from the MK metric by a (finite) conformal
transformation (and subsequent radial coordinate transformation) if
$\Omega(r) = \vpsi(r)/\vpsi_0$, where $\vpsi(r)$ is given by
(\ref{eq:phisoln}) with $a$ arbitrary. This provides some
insight into previous work seeking to use gauge transformations in CG
to make matter attractive to null
geodesics in the MK metric irrespective of the sign of its linear
term \cite{Edery01,Sultana17}.
In particular, setting $a=1/\gamma$ in (\ref{eq:phisoln}) and
performing a joint conformal and (radial) coordinate transformation as
outlined in Section~\ref{sec:rotcurvesEinstein}, one finds on
neglecting any products of $\beta$ and/or $\gamma$ that
$\tilde{B}'(r') \approx 1-2\beta/r' - \gamma r' - kr^{\prime 2}$,
which has the same form as the MK metric function (\ref{eqn:MKsoln})
at this level of approximation, but with the sign of the linear term
reversed. Nonetheless, this result must be treated with
some caution, since one finds that the exact expression for
$\tilde{B}'(r')$, without making any such approximations, does {\it
  not} have the same form as (\ref{eqn:MKsoln}) with merely a linear
term of opposite sign.

\section{Conclusions}
\label{sec:conc}

We have revisited the most celebrated phenomenological consequence of
CG, namely that the MK vacuum solution for a static,
spherically-symmetric system predicts flat galaxy rotation curves,
without the need for dark matter
\cite{Mannheim89,Mannheim93,Mannheim97,Mannheim11b,Mannheim12b,Mannheim17,OBrien18}. This
prediction is based on the assumption that massive (test) particles
have fixed rest masses and follow timelike geodesics in the MK
metric. The conformal equivalence of the MK and SdS metrics raises
concerns, however, that this prediction may be a gauge artefact, since
performing a similar analysis in the SdS metric yields rotation curves
without any flat region, as is well known. Since CG is (by construction) invariant to such transformations, they
should not change the observable consequences of the theory, unless
the conformal symmetry is broken in some way, either dynamically or by
imposing boundary conditions.  Indeed, if boundary conditions
are involved, interesting physics can arise quite generally from
differences between solutions that are gauge transformations of each
other, an obvious example being the Aharanov--Bohm effect
\cite{ABeffect}. Moreover, some care must clearly be taken regarding boundary
conditions at infinity for both the MK and SdS metric, since neither
is asymptotically flat, although the presence of both a constant term
differing from unity and a linear term in the MK metric exascerbates
this difficulty relative to the SdS metric. It is therefore
interesting that imposing a simple Neumann boundary condition on the
metric selects the vacuum solutions of GR with a cosmological constant
from the wider set of vacuum solutions of CG \cite{Maldacena11}, and
hence selects the SdS metric rather than the MK metric for static,
spherically-symmetric systems.

If the conformal symmetry is unbroken, however, the key to
resolving the question of which rotation curves are physically
realised is to recognise that massive (test) particles constitute some
form of `ordinary' matter, typically represented by a Dirac field,
which must generate its mass dynamically through interaction with a
scalar field. The consequent necessity for a scalar field that is
non-zero everywhere means that, in general, the spacetime outside a
static, spherically-symmetric matter source in CG is not described by
the MK vacuum solution, as demonstrated in
\cite{Brihaye09}. Nonetheless, a special solution is identified in
\cite{Brihaye09} for which the metric retains the MK form, since the
scalar field energy-momentum vanishes, despite the field being
non-zero everywhere. Indeed, such `ghost solutions' are
found in other physical contexts \cite{Davis75,Letelier75,Dimakis85}.

Despite having no effect on the geometry, ghost solutions
  can have important dynamical effects, and so are not `trivial', as
  claimed in \cite{Sultana17}. This is especially true for the scalar
field in the special solution obtained in \cite{Brihaye09}, since it
facilitates dynamical mass generation through its interaction with the
Dirac field that represents `ordinary' matter. Since the scalar field
is radially dependent in the MK frame, so too are the
masses of test particles, which therefore do not follow timelike
geodesics. In particular, on making a conformal transformation to the
Einstein gauge, in which the scalar field takes a constant value
everywhere and so massive particles do follow timelike geodesics, the
resulting metric is equivalent merely to the standard SdS form, which
lacks the linear term in the MK metric that is key to the successful
fitting of flat galaxy rotation curves \cite{Horne16}. Moreover, by
considering directly the motion of a Dirac particle in the presence of
a non-uniform scalar field, we further show that massive particles in
the MK frame also follow timelike geodesics of the SdS metric, as they
must for conformal invariance to hold. This therefore resolves the
apparent dependence of the physical predictions on the frame in which
the calculation is performed. More importantly, this unambiguously
identifies the rotation curves of the SdS metric, rather than the MK
metric, as those that are physically realised, which have no flat
region and hence do not fit observations of galaxies. We further show that the scalar field equation of motion introduces
  an additional constraint relative to the vacuum case, such that the
  coefficient of the quadratic term in the SdS metric is interpreted
  most naturally as proportional to a global cosmological constant;
  this therefore also precludes the modelling of rising rotation
  curves by fitting this coefficient separately for each galaxy, as
  performed in \cite{Mirabotalebi08}, albeit in the context of
  Weyl--Dirac gravity.

In addition, independently of the considerations of \cite{Brihaye09},
we show that the general form of the conformal transformation linking
MK and Einstein frames is picked out uniquely as that which preserves
the general form of any diagonal static, spherically-symmetric metric
of the form (\ref{eqn:MKform}), namely with a radial coefficient that
is (minus) the reciprocal of its temporal one.

Finally, we briefly discussed the consequences of our analysis for the
study of gravitational lensing in the MK metric, which has caused
considerable confusion in the literature, with many analyses producing
contradictory results
\cite{Edery98,Pireaux04a,Pireaux04b,Amore06,Sultana10,Sultana12,Villanueva13,Cattani13,Lim17,Oguzhan19,Turner20}.
One may straightforwardly resolve these disagreements by instead performing
calculations in the SdS frame, for which previous uncertainties
regarding the effects of the `cosmological constant' term have now
been clarified \cite{Butcher16}. One can then perform a conformal
transformation and accompanying radial coordinate transformation to
the MK frame, if desired. We also comment on the limited validity of
previous work seeking to make matter attractive to null geodesics in
the MK metric, irrespective of the sign of the linear term, by an
appropriate choice of conformal gauge \cite{Edery01,Sultana17}

\begin{acknowledgments}
We thank Philip Mannheim for his comments on the original version of
this paper, and the anonymous referee for several useful suggestions.
\end{acknowledgments}

\end{document}